# Enhanced spin injection in molecularly functionalized graphene via ultra-thin oxide barriers

J.C. Toscano-Figueroa[1,2,a)], N. Natera-Cordero[1,2,a)], D. A. Bandurin[1,b)], C.R. Anderson[1], V.H. Guarochico-Moreira[1,3], I.V. Grigorieva[1,c)], I.J. Vera-Marun[1,c)]

[1]Department of Physics and Astronomy, University of Manchester, United Kingdom
[2]Consejo Nacional de Ciencia y Tecnología (CONACyT), México
[3]Departamento de Física, Escuela Superior Politécnica del Litoral, Ecuador

a) J. Toscano, N. Natera and D. Bandurin contributed equally to this work.
b) Present address: Department of Physics, Massachusetts Institute of Technology, Cambridge, Massachusetts 02139, USA.
c) Authors to whom correspondence should be addressed:
Irina.V.Grigorieva@manchester.ac.uk and Ivan.VeraMarun@manchester.ac.uk

*Realisation of practical spintronic devices relies on the ability to create and detect pure spin currents. In graphene-based spin valves this is usually achieved by injection of spin-polarized electrons from ferromagnetic contacts via a tunnel barrier, with $Al_2O_3$ and MgO used most widely as barrier materials. However, the requirement to make these barriers sufficiently thin often leads to pinholes and low contact resistances which in turn results in low spin injection efficiencies, typically 5% at room temperature, due to the so-called resistance mismatch problem. Here we demonstrate an alternative approach to fabricate ultra-thin tunnel barrier contacts to graphene. We show that laser-assisted chemical functionalization of graphene with $sp^3$-bonded phenyl groups effectively provides a seed layer for growth of ultrathin $Al_2O_3$ films, ensuring smooth, high quality tunnel barriers and an enhanced spin injection efficiency. Importantly, the effect of functionalization on spin transport in the graphene channel itself is relatively weak, so that the enhanced spin injection dominates and leads to an order of magnitude increase in spin signals. Furthermore, spatial control of functionalization using a focused laser beam and lithographic techniques can in principle be used to limit functionalization to contact areas only, further reducing the effect on the graphene channel. Our results open a new route towards circumventing the resistance mismatch problem in graphene-based spintronic devices based on the easily available and highly stable $Al_2O_3$, and facilitate a step forward in the development of their practical applications.*

Interest in low-dimensional spintronics has intensified during the last decade following the demonstration of graphene as a truly two-dimensional spin channel [1] exhibiting long spin relaxation lengths at room temperature [2,3]. Graphene also offers gate-tunable carrier concentration and high mobility, which makes it ideal for the creation of spin-based electronic devices, such as transistors and logic gates [4,5]. An essential part of such devices is a tunnel barrier for spin injection that ideally should be uniform and atomically thin, in order to achieve an optimum combination of a sufficiently high resistance and a high enough tunneling current [1,6]. Traditionally used tunnel barriers are based on thin oxide films, such as $Al_2O_3$ and MgO [1,6,7]. These are grown via electron beam evaporation and typically exhibit spin injection efficiencies at room temperature of ≈5% [3,8]. Similar efficiencies were obtained for $Al_2O_3$ grown via atomic layer deposition [9]. One of the reasons for such relatively low spin injection efficiency is the presence of pinholes, as they cause direct ohmic contact and a large suppression of spin polarization due to a fundamental limitation known as the resistance mismatch [10]. The latter limits spin injection efficiency when the (spin) resistance of an injector contact, $R_c$, is lower than the typically high spin resistance of the graphene channel, $R_{ch}^s = \rho \lambda_s / W$ (here $\rho$ is the sheet resistance of graphene, $W$ the width of the channel, and $\lambda_s$ its spin relaxation length) [2]. This results in contact-induced spin relaxation, which greatly decreases the measured spin signals and reduces the spin relaxation time [11].

Therefore, it is essential to limit formation of pinholes in ultra-thin tunnel barriers. Sputter deposition, a standard industrial method to grow barriers for tunnel magnetoresistance structures, was shown to result in a nearly pinhole free ≈1 nm thick $Al_2O_3$ barrier on graphene [12,13]. However, this method destroyed the structural integrity of the graphene channel, even leading to its amorphization in the case of MgO deposition. More successful was an alternative approach which involved fabrication of MgO barriers using a Ti seed layer and making the contacts vary narrow (~50 nm wide); this yielded a spin injection efficiency of up to 30% [6]. Another method involved initial evaporation of Co/MgO electrodes on a $SiO_2$ substrate, followed by mechanical transfer of a graphene-hBN stack on top of these electrodes, leading to suppression of contact-induced spin relaxation [14]. However, no similar successes have been reported for $Al_2O_3$ barriers. Other efforts to increase the polarization of injected electrons in graphene include using barriers based on functionalized graphene with fluorine [15], amorphous carbon [16], and strontium oxide [17]. Two-dimensional crystals, such as hBN, can act as ideal ultrathin tunnel barriers themselves [18] and have been used to demonstrate efficient spin injection in graphene devices [19,20]. Nevertheless, practical implementation of the latter approach depends on overcoming the formidable challenge of scaling up the identification and mechanical transfer of such ultrathin hBN layers.

Here we show that the problem of resistance mismatch can be circumvented by functionalization of designated graphene regions using a photo-activated chemical reaction and subsequent deposition of oxide tunnel contacts. To this end, we demonstrate a controlled and reproducible introduction of $sp^3$ defects in monolayer graphene using chemical reaction with benzoyl peroxide (BPO). The reaction can be easily and reliably controlled by irradiation with laser light, with defect density determined by the exposure. Functionalization introduces topographic roughness of the graphene surface, which in turn leads to a more uniform formation of the insulating tunnel barriers. The resulting devices are characterized by an increased spin tunnelling polarization without a significant suppression of the spin transport parameters in the channel. Our approach offers a convenient and easily scalable method for the development of graphene spintronics with improved spin injection efficiency.

To achieve functionalization, we follow the process reported in ref. [21] where photochemical reaction between BPO solution and graphene was used to introduce spatially localized defects into the basal plane of graphene. As demonstrated previously, on contact with the solution, BPO molecules efficiently physisorb on the surface of graphene [22], and can subsequently be broken up to form $sp^3$ bonds with graphene on exposure to laser light [21,23]. The BPO molecules consist of two benzene rings connected by an oxygen bridge; laser radiation breaks up the bridge, producing highly reactive phenyl radicals that covalently bond to graphene, effectively playing the role of 'adatoms' as each radical bonds with just one carbon atom [21]. To initiate the functionalization reaction, we used focused laser beam (532 nm wavelength). The extent of functionalization (the average density of covalently attached phenyl radicals) was controlled by varying the radiation power, $P_R$, and the exposure time (See Supplemental Material for further details).

Figure 1b shows an optical photograph of a typical graphene flake which was processed by the protocol described above. After the reaction, the flake exhibited no visible changes in optical microscopy but Raman spectroscopy revealed a modification of the crystal structure apparent from the rise and evolution of D and D' bands upon increasing the exposure during functionalization (Fig. 1d). The emergence of these Raman bands indicates the presence of point defects in the graphene lattice, while their relative intensity ratio ($I_D/I_{D'}$) can serve as a tool to probe the defect type. Previous studies [24] established that the $sp^3$-hybridization of otherwise flat $sp^2$ graphene bonds results in $I_D/I_{D'}$~13, a value close to that found in our experiments. This is consistent with the expected $sp^3$-character of the C-C bonds between graphene and phenyl radicals [21]. Similarly, the ratio of D and G bands ($I_D/I_G$) in the Raman spectra provides information on the average density of the introduced defects [24]. We found that $I_D/I_G$ can be controlled within a range of 0—2 by varying the exposure during functionalization, as is apparent from the $I_D/I_G$ ratio mapped onto differently exposed areas of the flake, Fig. 1b: different regions correspond to different radiation power densities from the laser beam, $P_R$. For the areas with most exposure, the $I_D/I_G$ ratio exceeds unity (see inset of Fig. 1d). For example, a radiation power density of $P_R$ ~5 mW/μm$^2$ translates to high functionalization and ~0.01% defect coverage, ~5x10$^{11}$ defects/cm$^2$ (see ref. [24] for details of calculating the defect density from Raman spectra).

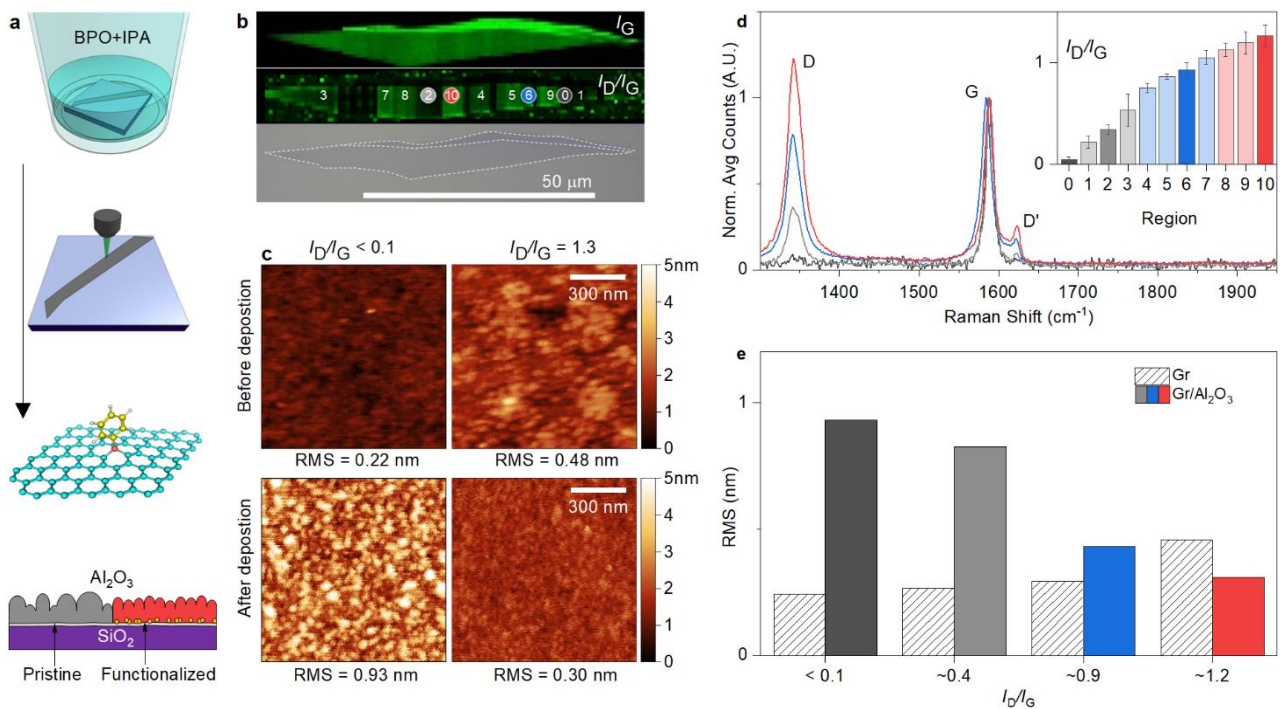

**Fig. 1. Controlled functionalization of monolayer graphene and its effect on the growth of an Al$_2$O$_3$ barrier. a,** Schematic representation of the functionalization process (top to bottom), with the resulting Al growth morphology (bottom). **b,** Raman maps of the G peak (top), D/G intensity ratio (middle), and the corresponding optical image of the graphene flake (bottom). Different regions correspond to different laser irradiation doses. The larger outlined area in the optical image is the monolayer. **d,** Typical Raman spectra corresponding to regions 0 (black), 2 (grey), 6 (blue) and 10 (red) in **b**. Inset: D/G ratio for all functionalised regions in (b). **c,** AFM images of the areas of graphene with lowest (region 0) and highest (region 10) densities of $sp^3$ defects (top row) and of the same areas after deposition and oxidation of Al (bottom row). **e,** Roughness of the oxide film on top of graphene areas with different degrees of functionalization (coloured bars) and of graphene prior to Al deposition (hatched bars). The degree of functionalization is indicated by the D/G ratio.

To verify that the laser-assisted BPO-graphene reaction introduces only adatom-type $sp^3$ defects (rather than, for example, carbon vacancies) and is reversible, we have performed additional experiments where functionalized graphene flakes were annealed at 300°C for 3 hours in inert Ar atmosphere. This led to a strong suppression of the D and D' peaks in the Raman spectra as shown in Fig. S1, i.e., to the removal of the defects. We note that this is only possible for adatoms as the used annealing temperature is too low to 'heal' vacancies [25] (see Supplemental Material for further details).

For further analysis we examined the graphene surface using atomic force microscopy (AFM) and analysed the flakes' roughness. The latter is usually quantified by a root-mean-square (RMS) parameter which provides a measure of the average height variation of the sample under study. As shown in Fig. 1c, we find that non-functionalized graphene areas exhibited small RMS values, typically ~0.2 nm, which is comparable to that of the underlying $SiO_2$. In contrast, functionalized regions were characterized by a much larger RMS ~0.5 nm pointing to their morphological modification by the $sp^3$ bonded phenyl radicals.

For the purpose of the present study, after functionalization we deposited 0.6 nm of aluminium and allowed its complete oxidation in a controlled $O_2$ environment. After the deposition, we repeated the RMS analysis and found that the $Al_2O_3$ layer formed over functionalized graphene areas was much smoother, with RMS of only 0.3 nm. This is to be compared with the $Al_2O_3$ barrier grown on pristine graphene areas that exhibited more than 3 times higher RMS. This proved to be a consistent trend when analysing different degrees of functionalization, as shown in Fig. 1e. Although the finding that rougher graphene leads to smoother $Al_2O_3$ films may seem counterintuitive, it is in fact in agreement with expectations if we recall that deposited metal atoms are known to be highly mobile on pristine graphene [6,26] which leads to significant non-uniformities and/or roughness of ultrathin films. Making graphene less smooth suppresses the high surface diffusion of the metal, resulting in smooth, uniform films. Further details of our surface roughness analysis are given in the Supplemental Material.

An improved uniformity of the tunnel barriers turned out to be the key parameter ensuring efficient spin-injection, as we now proceed to demonstrate. To this end, we fabricated several graphene spin-valves having several regions of different functionalization extent. Ferromagnetic Co electrodes were deposited on top of $Al_2O_3$, to form magnetic tunnel contacts. Each functionalized area was endowed with at least a pair of tunnel contacts (Fig. 2a) which allowed systematic analysis. Three-terminal measurements of the junctions' contact resistance, $R_c$, showed that it is always greater in the functionalized areas as compared to pristine (unexposed to laser light) regions (Fig. 2c). For example, the product of $R_c$ and the contact area, $A$, reaches 100 kΩ/μm² in the heavily functionalized tunnel junctions, a more that 100 times increase compared to $Al_2O_3$/Co contacts grown on pristine graphene where $R_c A$ is typically below 1 kΩ/μm².

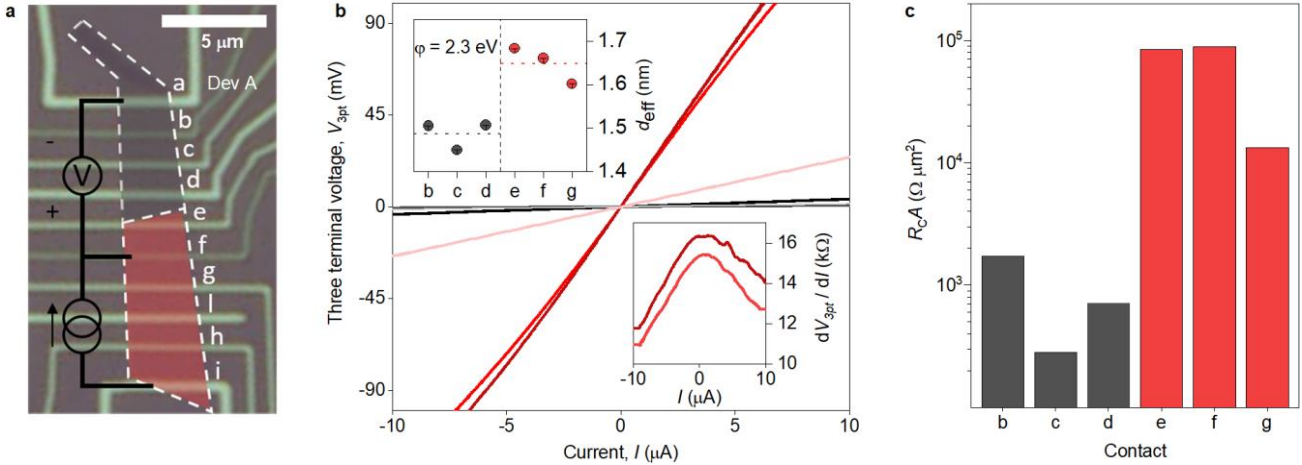

**Fig. 2. Effect of functionalization on contact resistance. a,** Optical image of device A; the dashed line delimits the regions of the graphene channel that were exposed to laser light at high power density (bottom part) and only exposed to BPO but not to laser light (top part). Corresponding $I_D/I_G$ ratios are 0.1 and 1.9. **b,** I-V curves for contacts e—g in the highly functionalized region (red curves) and b—d in the unexposed region (grey curves) measured in a three-terminal configuration shown in (a). *Left inset*: effective tunnel barrier thickness estimated using the Simmons model (Supplemental Material). *Right inset*: differential resistance for contacts e & f, clearly showing nonlinear dependence on the current. **c,** Resistance-area product for contacts b—g. Bars are coloured red for contacts in the highly functionalized region and dark grey for unfunctionalized graphene.

Consistent with the high contact resistance for heavily functionalized areas of graphene, the three-terminal I-V curves exhibited a clear non-linear behaviour shown in Fig. 2b. This nonlinearity is more apparent on the derivative plots of the I-V curves, which reveal a clear signature of the tunnelling behaviour: d$V$/d$I$ decreases with increasing $I$. In contrast, I-V curves measured on the pristine graphene regions remained linear (ohmic) over the entire $I$-range probed, suggesting the presence of pinholes in the $Al_2O_3$ barrier. Using Simmons model [27] to fit the measured I-V curves for heavily functionalized areas, we were able to estimate an effective tunnel barrier thickness $d_{eff}$ and the height of the energy barrier, $\varphi$=2.3 eV (the latter is in good agreement with literature [28]). The above $\varphi$ value was then used to determine the effective barrier thickness for linear I-Vs, such as black and grey curves in Fig. 2b. See the Supplemental Material for further details of our fitting procedure. The inset of Fig. 2b compares $d_{eff}$ for the defect-seeded tunnel junctions and pristine areas: there is a clear increase of $d_{eff}$ in the functionalized areas, particularly for the larger $R_cA$ (contacts e and f).

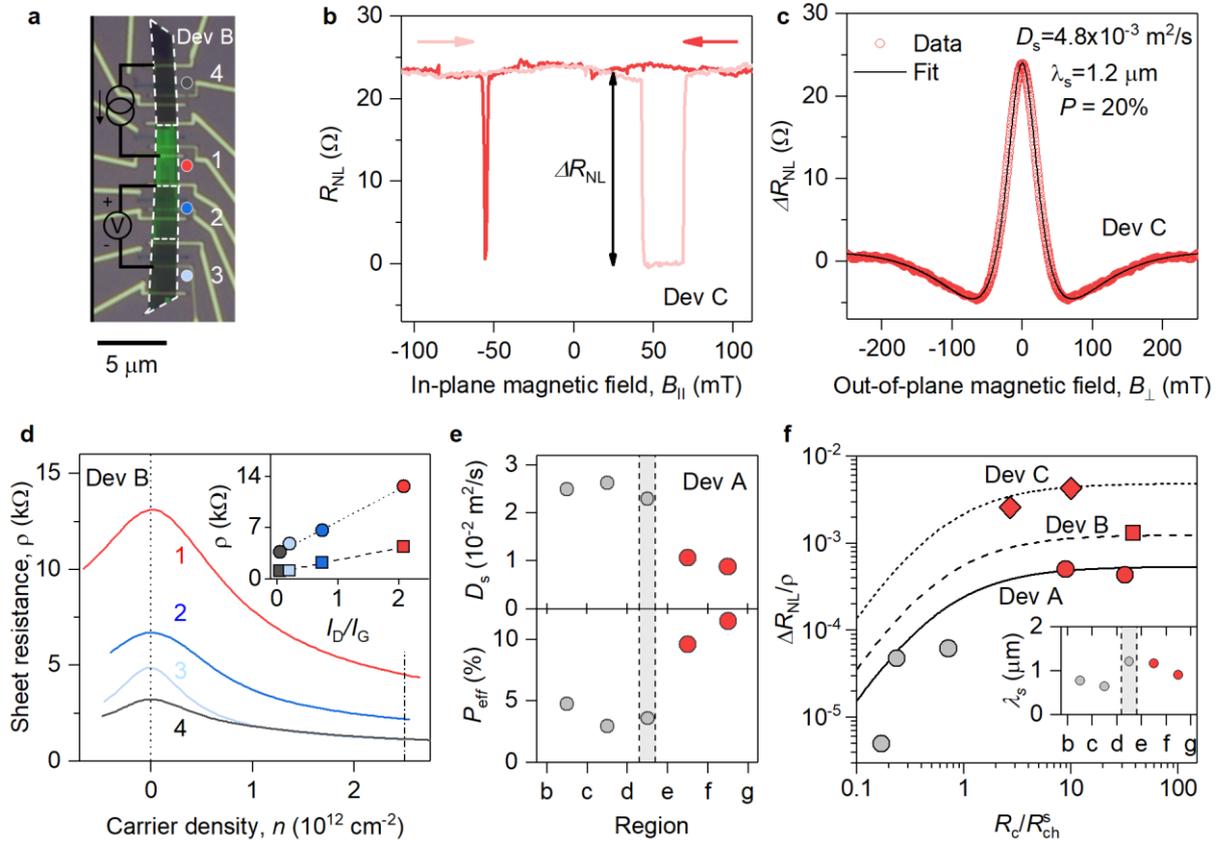

**Fig. 3. Spin transport in functionalized graphene and resistance mismatch. a,** Optical image of device B, with four differently functionalized regions (1—4), indicated by the white-dashed line and a superimposed Raman map showing D/G intensity ratio (ranging from 0 to 2.1). **b,** Example of the non-local response measured across a highly functionalized area with $I_D/I_G$ = 1.1 and $P$ = 20% in a different device (C). **c,** Spin precession (Hanle curve) for the same area as in (b). **d,** Resistance vs carrier density for the different regions of device B shown in (a). Inset: sheet resistance for the same regions at the neutrality point (circles) and at $2.5\times10^{12}$ cm$^{-2}$ (squares) as a function of the $I_D/I_G$ ratio. **e,** Spin diffusion coefficient and contact polarization obtained from spin-transport experiments for different regions of device A shown in Fig. 2. The $I_D/I_G$ ratios are 0.1 (1.9) for the grey (red) circle symbols. **f,** Scaled non-local spin-valve signal vs the ratio of contact resistance to spin channel resistance, with fits to the Takahashi-Maekawa model (see text). Circular symbols correspond to the same regions from device A as in (e), with the continuous line being the model for $P$ = 10%. Square symbol corresponds to region 1 from device B in (a), with $I_D/I_G$ = 2.1 and dashed model line with $P$ = 9%. Diamond symbols correspond to two regions from another device (C, not shown), with $I_D/I_G$ = 1.1 and dotted model line with $P$ = 19%. Inset: spin relaxation length for the same regions as in (e).

For further analysis, we performed standard measurements of the non-local resistance $R_{NL}$ as a function of in-plane magnetic field, $B_{\parallel}$ (see Fig. 3a for the measurement configuration). This showed a pronounced switching behaviour, consistent with the magnetisation reversal of the magnetic electrodes [3,5] and propagation of pure spin current along the functionalized graphene areas, suggesting a weak effect of functionalization on the spin transport. Furthermore, the change in the non-local signal, $\Delta R_{NL}$, between parallel and anti-parallel magnetization states reaches values in the range of 20—30 Ω, which is at least an order of magnitude larger than typical values for tunnel junctions grown on pristine graphene and is comparable to the best graphene spin valves reported so far [6]. The width of the channel in our devices is in the range of 1—3 μm. Note that wider devices, e.g., using CVD graphene, with similar polarizations would exhibit correspondingly smaller signals. Changing the direction of the applied magnetic field to be perpendicular to the graphene channel results in

Larmor spin precession and further dephasing, also known as a Hanle curve. This leads to a sign reversal of $R_{NL}$ vs $B_\perp$ when the electronic spin accumulation undergoes a π precession (Fig. 3c). Fitting the measured $R_{NL}(B_\perp)$ to a standard solution of the Bloch equation allowed us to extract the spin diffusion properties of the device: spin relaxation length, $\lambda_s$, spin diffusion coefficient $D_s$, and effective spin polarization, $P_{eff}$. Figure 3e shows the latter two characteristics for different regions within the same device: pristine (i.e., unexposed to laser light) with $I_D/I_G < 0.1$, and heavily functionalized, $I_D/I_G = 1.9$. As one can see from this direct comparison within a single graphene flake, heavily functionalized regions demonstrate a substantial increase of $P_{eff}$, up to threefold, compared to the unfunctionalized areas, pointing to an improved spin injection efficiency of the defect-seeded tunnel junctions. Functionalization also leads to an overall enhanced performance of the spin-valve devices. This is evident in Fig. 3f, where the non-local spin signal for regions with high functionalization is more than an order of magnitude larger than for unfunctionalized areas of the same device.

To elucidate the role of the magnetic tunnel contacts, Fig. 3f presents the spin signal as a function of the ratio of the contact resistance, $R_c$, to the spin resistance of the graphene channel, $R_{ch}^S$. Within the well-established model by Takahashi and Maekawa [29], both of these resistances act as parallel channels for spin transport. Therefore, $R_c/R_{ch}^S$ represents the ratio of injected spins that continue diffusing and eventually relax inside the graphene channel, to those spins that are reabsorbed by the magnetic electrode, where they rapidly relax, making this ratio a good figure of merit for the effect of relaxation induced by invasive contacts [11,30]. When $R_c/R_{ch}^S < 1$, spin transport is considered to occur in a regime of spin resistance mismatch, where the effective spin injection efficiency of the magnetic tunnel contacts is reduced. In our devices, graphene regions with low functionalization (or not functionalized at all) are within this regime. On the other hand, for highly functionalized graphene, $R_c/R_{ch}^S \gg 1$, resistance mismatch is minimized, leading to a large increase in $\Delta R_{NL}$. This difference between unfunctionalized and functionalized regions was found to be consistent and reproducible across all studied devices: the former showing $P_{eff} \leq 5\%$ and the latter $P_{eff}$ = 10—20%. Within the functionalized regions, there was no clear correlation between $P_{eff}$ and the degree of functionalization but it should be noted that the number of studied devices is too small for statistical analysis.

The observed increase in spin signals for our functionalized devices can in principle originate from a different scenario, other than the improvement of the spin injection efficiency of the contacts. Even in a regime free of spin resistance mismatch, the spin signal depends on several parameters, as per the expression: $\Delta R_{NL} \propto P^2 \rho \lambda_s e^{-L/\lambda_s}/W$ [29]. In our case the spin relaxation length was found to change only weakly with an increasing degree of functionalization, with $\lambda_s$ in the range between 1 and 1.5 μm for all our devices (see inset of Fig. 3f). Therefore, the increased spin signal cannot be explained by a change in spin relaxation length. This observation is in agreement with previous work on graphene functionalized with chemisorbed hydrogen, which showed that hydrogenation had little effect on the spin relaxation length [31], whereas another work found a suppression of the spin relaxation length [32]. Unlike the spin relaxation length, the sheet resistance of graphene is known to be strongly affected by sp$^3$ defects as they introduce strong point-like potentials which cause significant momentum relaxation. The resulting increase in the sheet resistance should lead to a trivial increase of the spin signal via its linear scaling with $\rho$ [29]. Indeed, we observe such a monotonic increase in $\rho$ with the functionalization extent (see Fig. 3d) where highly functionalized graphene, with $I_D/I_G \sim 2$, exhibits a threefold increase in sheet resistance. Note that this observation from charge transport measurements is also consistent with the threefold decrease in spin diffusion coefficient shown in Fig. 3e, where $D_s$ was obtained from spin transport measurements on another device (device A, shown in Fig. 2a) with a similar level of functionalization ($I_D/I_G$ = 1.9). Such inverse scaling between $D_s$ and $\rho$ is indeed expected for spin transport carried by electrons, in the absence of any dominant electron-electron interactions, e.g. spin Coulomb drag [32,33].

To account for this trivial $\Delta R_{NL} \propto \rho$ scaling, Fig. 3f shows $\Delta R_{NL}$ normalised to the sheet resistance of each corresponding graphene region. This allowed us to remove the contribution due to the increased momentum scattering in functionalized graphene and focus exclusively on the role of the spin injection and detection efficiency of the contacts. Following Takahashi and Maekawa [29], we have modelled the spin signal for functionalised regions of several devices, where we used the corresponding sheet resistances and spin relaxation lengths for each region. The results in Fig. 3f show that devices with highly functionalized regions exhibit contact polarizations at room temperature in the range of 10—20 %. For one of the devices (device A) we were able to compare our data with the Takahashi-Maekawa model for the entire range of $R_c/R_{ch}^s$ (solid line in Fig. 3f) which showed a good agreement for $P \sim 10\%$ for the intrinsic spin polarisation of the contacts (here the $P$ value corresponds to the effective polarisation in Fig. 3e). These results provide a clear evidence that the notable enhancement of spin signals in our devices is indeed a consequence of the higher contact resistance and effective polarization $P_{eff}$.

In its turn, the increased contact resistance is clearly the result of the much reduced roughness and the absence of pinholes in our ultrathin $Al_2O_3$ tunnel barriers. We speculate that the reduced roughness is an indication of finer grains in these polycrystalline films, as the grain size is known to strongly depend on the growth conditions (see e.g. refs. [34,35]). To this end, we further analysed the surface morphology in AFM images of the $Al_2O_3$ films deposited on graphene with and without high levels of functionalization, i.e., with $I_D/I_G$ ratios of <0.1 and 1.3 respectively (see Supplemental Material for details). This showed that the height distribution in the AFM images for $Al_2O_3$ films deposited on highly functionalized graphene are significantly narrower than for pristine regions, consistent with the extracted RMS values. Similarly, the distribution of the lateral sizes of the 'grains' in the AFM images for highly functionalized graphene is peaked more sharply and at a lower size compared to the $Al_2O_3$ film grown on unfunctionalized graphene. This indicates that $sp^3$ bonded phenyl groups serve as nucleation centres for Al growth, resulting in smoother oxide films.

The nearly constant spin relaxation length (inset in Fig. 3f) and the decreased diffusion coefficient in our functionalized devices (Fig. 3e) imply that the spin lifetime increases as the result of functionalization. This observation is in agreement with previous work on spin transport in graphene functionalized with hydrogen [31] which also reported an enhancement of spin lifetime. As a possible explanation, it has been suggested [31] that the increased spin lifetime is consistent with the Dyakonov-Perel (DP) spin relaxation mechanism: An enhanced scattering by defects in our experiments is indicated by a notable increase of graphene resistance (Fig. 3d). We note that the weak dependence of the spin relaxation length on the presence of $sp^3$ defects is in contrast to predictions of ref. [36] where it was suggested that $sp^3$ defects should lead to a locally enhanced spin-orbit coupling in graphene, an enhanced spin-flip scattering at defects and a decrease of the spin relaxation length. However, this prediction did not consider other possible sources of spin-orbit coupling such as the substrate [37]. In our experiment, the dominant role of defects appears to be related to momentum scattering.

Finally, we note that the proposed laser-assisted BPO functionalization allows using thermal evaporation as a deposition technique to achieve tunnel barriers on graphene with both large $R_cA$ product and high spin polarization. Evaporation proved to be beneficial for fabrication of spintronic devices as it avoids damage to graphene during deposition, as opposed to sputtering that can also create barriers with large $R_cA$ product [12] but at the expense of a large defect density in the graphene channel [13]. Furthermore, the possibility to control the amount of sp$^3$ defects in graphene independently of the barrier deposition process creates possibilities to control spin transport utilising the magnetic nature of sp$^3$ defects [32,38,39].

We acknowledge support from the European Union's Horizon 2020 research and innovation program under Grant Agreement Nos. 696656 and 785219 (Graphene Flagship Core 2) and from the Engineering and Physical Sciences Research Council (UK) EPSRC CDT Graphene NOWNANO EP/L01548X. J.C.T.-F. and N.N-C. acknowledge support from the Consejo Nacional de Ciencia y Tecnología (México). D.A.B. acknowledges support from the FP7 Marie Curie Initial Training Network "Spintronics in Graphene" (SPINOGRAPH). C.R.A. acknowledges support from the EPSRC Doctoral Training Partnership (DTP). V.H.G.-M. acknowledges support from the Secretaría Nacional de Educación Superior, Ciencia y Tecnología (Ecuador). I.J.V.M. acknowledges support from the FP7 FET-Open Grant 618083 (CNTQC). Research data are available from the authors upon request.

# Enhanced spin injection in molecularly functionalized graphene via ultra-thin oxide barriers

J.C. Toscano-Figueroa[1,2,a)], N. Natera-Cordero[1,2,a)], D. A. Bandurin[1,b)], C.R. Anderson[1], V.H. Guarochico-Moreira[1,3], I.V. Grigorieva[1,c)], I.J. Vera-Marun[1,c)]

[1]Department of Physics and Astronomy, University of Manchester, United Kingdom
[2]Consejo Nacional de Ciencia y Tecnología (CONACyT), México
[3]Departamento de Física, Escuela Superior Politécnica del Litoral, Ecuador

a) J. Toscano, N. Natera and D. Bandurin contributed equally to this work.
b) Present address: Department of Physics, Massachusetts Institute of Technology, Cambridge, Massachusetts 02139, USA.
c) Authors to whom correspondence should be addressed:
Irina.V.Grigorieva@manchester.ac.uk and Ivan.VeraMarun@manchester.ac.uk

## Supplemental Material

### Functionalization

Preparation of functionalized graphene involved several steps. First, graphene flakes were deposited on top of an oxidized Si substrate (290 nm of $SiO_2$) using standard mechanical exfoliation of highly oriented pyrolytic graphite. Monolayer graphene was identified by optical microscopy and Raman spectroscopy. The substrate was then immersed in a benzoyl peroxide (BPO) / isopropyl alcohol (IPA) solution (7 mg / 25mL) for 15 min, resulting in physisorption of BPO molecules on graphene [1]. Next, the substrate was taken out of the solution and the solution excess was removed by $N_2$ blow-dry. To initiate the functionalization reaction [2,3], we used the built-in laser of the Renishaw Raman spectrometer, with a focused laser beam of wavelength 532 nm, which allowed accurate control of the radiation power, $P_R$, and the radiation time. These parameters were used to control the extent of the functionalization. We kept the focused laser beam to either 5% or 10% of the total available power (37.9 mW), leading to a typical radiation power density of $P_R$ ~5 mW/μm$^2$, and a radiation time for each laser position from 0.5 to 3 seconds. Finally, the processed graphene flakes were rinsed in IPA, blow-dried in $N_2$ and characterised using Raman spectroscopy. The achieved range of the intensity ratio $I_D/I_G$ from <0.1 to ~2 corresponds to defect densities between 1x10$^{10}$ defects/cm$^2$ and 5x10$^{11}$ defects/cm$^2$, or 0.0002% to 0.01% defect coverage.

### Reversibility of graphene functionalization

Besides the expected sp$^3$-bonded phenyl radicals, carbon vacancies are another type of defects that could contribute to the Raman and transport signatures observed in this work. To rule out any significant contribution from carbon vacancies, and therefore confirm that the defects correspond to sp$^3$ bonded phenyl, we have performed additional experiments studying the change of the ratio of D and G bands ($I_D/I_G$) after heat treatment. To this end we first functionalized graphene flakes as described above and then annealed them under an inert Ar atmosphere at 300°C for 3 hours. Figure S1 shows Raman spectra for monolayer graphene corresponding to its pristine state, after functionalization, and after the heat treatment. This control experiment demonstrated that the D band could be almost fully suppressed by heat treatment. Such reversibility of the functionalization process gives further evidence that the nature of the defects are not carbon vacancies, as these temperatures are too low to heal possible vacancies in graphene [4].

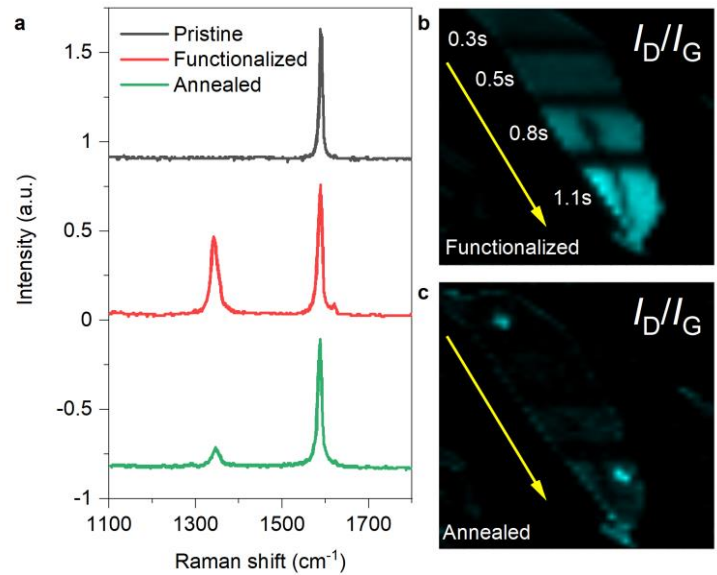

**Figure S1. Reversibility of the functionalization process in monolayer graphene.** (a) Raman spectra for pristine, functionalized, and annealed monolayer graphene, for an exposure time of 0.8 s. (b) Raman map of the $I_D/I_G$ ratio obtained on the functionalized monolayer graphene flake, including regions with different exposure times. (c) Similar Raman map as in (b) after the heat treatment.

## *Device fabrication*

Standard electron beam lithography was used for patterning of bi-layer resist coating applied to the substrates for fabrication of magnetic tunnel electrodes. Deposition of all layers was done using electron beam evaporation at a base pressure of ~$10^{-6}$ mbar. First, an Al layer with nominal thickness of 0.6 nm was deposited and subsequently oxidized in situ under a controlled $O_2$ pressure of $5\times10^{-3}$ mbar for 10 min, to create an oxide tunnel barrier. Next, we deposited 30 nm of Co for the magnetic electrode, followed by a capping layer of 5 nm of Al.

## *Electrical characterisation*

Electrical measurements were carried out in a high-vacuum environment (≤$10^{-5}$ mbar) of a cryostat operated at room temperature. Four-terminal measurements were performed using standard lock-in techniques, both to characterise the graphene resistivity and to perform non-local spin transport experiments. Three-terminal measurements, using DC biasing, were used to characterise the *IV* curves of the tunnelling contacts. For spin transport measurements we swept the value of an applied magnetic field, either in-plane, $B_{||}$, within ±150 mT for spin valve measurements, or out-of-plane, $B_\perp$, within ±400 mT for Hanle (spin precession) measurements.

## AFM analysis of surface morphology

The surface morphology in AFM images for graphene with and without high levels of functionalization, i.e., with $I_D/I_G$ ratios of <0.1 and 1.3 respectively, was performed in two ways. First, we extracted the height distributions for each functionalization level, both before and after deposition of the $Al_2O_3$ thin films. The resulting height distributions and their corresponding AFM images are shown in Fig. S2.

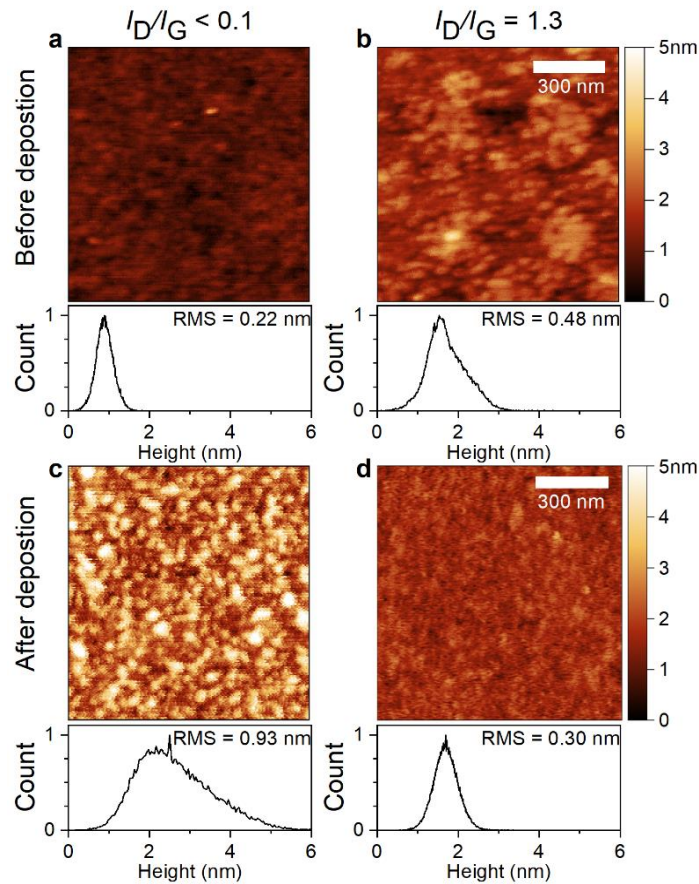

**Figure S2. AFM analysis of the surface morphology of graphene and $Al_2O_3$ films.** AFM images and corresponding height distributions for AFM images of the areas of unfunctionalized (a) and highly functionalized (b) graphene and of the same areas after deposition and oxidation of Al (c,d). Same AFM images as those shown in Fig. 1c.

As another measure of roughness, we analysed the distribution of lateral radii of surface bumps in the images of $Al_2O_3$ films. This analysis was performed using the watershed particle detection algorithm within the Gwyddion analysis software [5]. This is a classical image segmentation algorithm [6], where the AFM data are first inverted in the vertical direction and treated as a topographical map on which we place "virtual drops" that are left to relax to local minima, thus forming "lakes". The latter then represent individual 'bumps' in the AFM image, see Fig. S3. We have chosen this approach over simple threshold algorithms, as the latter strongly depend on a manually set global height threshold parameter and usually fail for rough surfaces or for agglomerated particles [7]. The use of the watershed algorithm ensured that the effect of substrate irregularities can be strongly suppressed. To this end, we first applied a pre-processing step consisting of a 2 pixel-Gaussian blurring to reduce single-pixel noise, then identified surface bumps using the watershed method. The bump size was extracted via an algorithm based on particle projection on the xy plane. Finally, the average bump radius, defined as the mean distance from the centre of mass of each bump to its boundary, was calculated to produce a normalized histogram for a given AFM image (see Fig. S3).

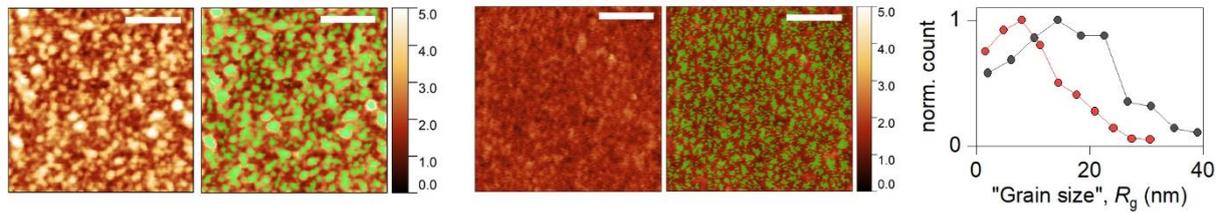

**Figure S3. Analysis of the lateral size of Al₂O₃ films' roughness.** Left: AFM image of the Al₂O₃ film grown on unfunctionalized graphene, and the same image with the surface 'bumps' identified via the watershed algorithm. Centre: same for the Al₂O₃ film grown on highly functionalized graphene. Right: resulting distributions for the unfunctionalized (grey symbols) and the highly functionalized (red symbols) cases. All scale bars are 300 nm.

## Roughness of SiO$_2$

For all studied devices, laser-irradiated areas were set to extend beyond the side edges of the corresponding graphene region. Therefore, the SiO$_2$ substrate adjacent to the graphene channel was also irradiated by the laser beam and considered to be also functionalised with phenyl radicals. Interestingly, similar trends in RMS were observed on SiO$_2$ to those seen on graphene (Fig. 1e in the main text) with regard to the level of functionalisation (here functionalization is quantified via Raman peaks for the adjacent graphene, as described in the main text).

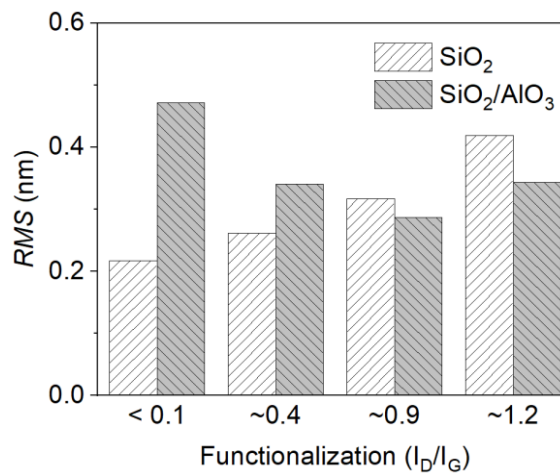

**Figure S4. Roughness of functionalized SiO$_2$ and of the Al$_2$O$_3$ film grown on it.** Shown are measured RMS on the SiO$_2$ substrate before (white hatched) and after (grey hatched) deposition of Al$_2$O$_3$ vs the degree of functionalization as determined for the adjacent graphene region (see text).

## Simmons model

The Simmons approximation [8] was used to analyse the *IV* curves of the tunnelling contacts.

$$I = A\left(\frac{e}{2\pi h d_{\text{eff}}^2}\right)\left\{\left(\varphi - \frac{eV}{2}\right)exp\left[-\frac{4\pi d_{\text{eff}}}{h}(2m)^{\frac{1}{2}}\left(\varphi - \frac{eV}{2}\right)^{\frac{1}{2}}\right] - \left(\varphi + \frac{eV}{2}\right)exp\left[-\frac{4\pi d_{\text{eff}}}{h}(2m)^{\frac{1}{2}}\left(\varphi + \frac{eV}{2}\right)^{\frac{1}{2}}\right]\right\}$$

Here $h$ is Planck's constant, $m$ and $e$ are the mass and charge of the electron and $V$ is the voltage across the tunnel barrier. This expression permits the extraction of both the height of the energy barrier ($\varphi$) and effective barrier thickness ($d_{\text{eff}}$) by fitting the measured current as a function of the applied bias. Both $\varphi$ and $d_{\text{eff}}$ values extracted from the fit should be treated as effective values, limited by the simple free-electron model considered. In our analysis they were only used to compare the characteristics of functionalised and non-functionalised graphene in the highly doped (metallic) regime, rather than taken as exact values. To perform the fitting, we first considered the contacts which clearly showed non-linear *IV* behaviour and fitted both parameters. The extracted $\varphi$ values, self-consistent within a 20% range, were then averaged to obtain a nominal barrier height for $Al_2O_3$ of 2.3 eV, consistent with commonly extracted values in the literature [9], validating our approach. This nominal $\varphi$ was then fixed and the fitting procedure repeated for the whole set of contacts, in order to extract the effective barrier thickness for each contact. The inset in Fig. 2b shows the results of this analysis, indicating that an effective *d* for the chemically-seeded tunnel junctions is 0.2 nm larger than for pristine areas. Such an increased effective barrier thickness is in agreement with smoother and more uniformly grown $Al_2O_3$ films on heavily functionalised graphene (see above).